\documentclass{aa}  

\usepackage{graphicx}
\usepackage{txfonts}
\begin{document}

   \title{New Circumstellar Structure in the T Tauri System}

   \subtitle{A NIR, High--Contrast Imaging Study}

   \author{M. Kasper
          \inst{1}
          \and
          K. K. R. Santhakumari
					\inst{2,3}
					\and
          T.M. Herbst
					\inst{2}
					\and
          R. K\"ohler
					\inst{4,5}
          }

   \institute{
	European Southern Observatory (ESO), Karl-Schwarzschild-Str.~2, 85748 Garching, Germany\\
	\email{mkasper@eso.org}
						\and
Max-Planck-Institute for Astronomy (MPIA), K\"onigstuhl~17, 69117 Heidelberg, Germany
		        \and
						International Max Planck Research School for Astronomy and Cosmic Physics at the University of Heidelberg, Germany
						\and
        Universit\"at Innsbruck, Institut f\"ur Astro- und Teilchenphysik, Technikerstr.~25/8, 6020 Innsbruck, Austria
						\and
        University of Vienna, Department of Astrophysics,
        T\"urkenschanzstr.~17 (Sternwarte), 1180 Vienna, Austria
             }

   \date{Received Month XX, XXXX; accepted Month XX, XXXX}

 
  \abstract
	{}
   {The immediate vicinity of T Tauri was observed with the new high-contrast imaging instrument SPHERE at the VLT to resolve remaining mysteries of the system, such as the putative small edge-on disk around T Tauri Sa, and the assignment of the complex outflow patterns to the individual stars. 
}
   {We used SPHERE IRDIS narrow-band classical imaging in Pa$\beta$, Br$\gamma$, and the $\nu$ = 1-0 S(1) line of H$_2$, as well as in the nearby continua to obtain high spatial resolution and high contrast images over the NIR spectral range. Line maps were created by subtracting the nearby continuum. We also re--analyzed coronagraphic data taken with SPHERE's integral field spectrograph in $J$- and $H$-band with the goal to obtain a precise extinction estimate to T Tauri Sb, and to verify the recently reported claim of another stellar or substellar object in the system. 
	}
   {A previously unknown coiling structure is observed southwest of the stars in reflected light, which points to the vicinity of T Tauri N. We map the circumbinary emission from T Tauri S in $J$- and $H$-band scattered light for the first time, showing a morphology which differs significantly from that observed in $K$-band. H$_2$ emission is found southwest of the stars, near the coiling structure. We also detect the H$_2$ emitting region T Tauri NW. The motion of T Tauri NW with respect to T Tauri N and S between previous images and our 2014 data, provides strong evidence that the Southeast-Northwest outflow triggering T Tauri NW is likely to be associated with T Tauri S. We further present accurate relative photometry of the stars, confirming that T Tauri Sa is brightening again. Our analysis rules out the presence of the recently proposed companion to T Tauri N with high confidence.
	}
   {}


   \maketitle
%

\section{Introduction}

T Tauri is the prototype of an entire class of young stellar objects and a complex system of at least three stellar components with multiple jets and outflows. It is located in the Taurus-Auriga star-forming region at a distance of $146.7\pm0.6$\,pc \citep{loinard07} with an age of 1-2 Myr \citep{kenyon95}. The optically visible primary, T Tauri N, is an early K star \citep{cohen79}, which has a companion at 0$\farcs$7 separation to the south found by speckle interferometry \citep{dyck82}. This infrared companion, T Tauri S, is very red, and its brightness fluctuates at all NIR and MIR wavelengths \citep{ghez91}. Using speckle holography, \citet{koresko00} found that T Tauri S is itself a close ($\approx$ 0$\farcs$1) binary, composed of the IR luminous T Tauri Sa and the early M-star companion T Tauri Sb. The masses of T Tauri Sa and Sb are $2.12\pm0.1M_\odot$ and $0.53\pm0.06M_\odot$, respectively, which indicates that T Tauri Sa is at least as massive as T Tauri N, despite the large contrast in visible light \citep{koehler15,schaefer14}. The orbital period for the Sa-Sb binary is 27$\pm$2 years with a semi-major axis of $12.5^{+0.6}_{-0.3}$\,AU, and the inclination of the system is $20^{+10}_{-6}$ degrees \citep{koehler15}. The orbit of the T Tauri S/N binary is not yet well constrained, with a minimum semi-major axis of 300\,AU and a period of at least 800 years. The true values may very well be considerably larger than this, based on the large range of orbital solutions still compatible with the data.

All three components are actively accreting and are as yet spatially unresolved, i.e. point-like, hydrogen recombination line emitters \citep[e.g.][]{kasper02, duchene05}. The near infrared brightness of T Tauri Sa is quite variable, due to changes in the accretion rate on timescales of days. Likely, the long-term variability is caused by a combination of accretion and extinction \citep{vanBoekel10} with a "redder when faint" character \citep{beck04}.  After a bright phase following the periastron of T Tauri Sb in 1995 with an apparent magnitude of up to $K\sim6.5$ in 2000, T Tauri Sa has faded in the last decade down to $K\sim10$ in 2005. Recently, T Tauri Sa became brighter again showing $K=7.7$ in our data of December 2014 (this paper).

The extinction toward T Tauri N was estimated by \citet{kenyon95} to be $A_V=1.39$\,mag. The extinction toward T Tauri Sb is much higher and changes only moderately around an average value $A_V=15$\,mag \citep{duchene05}. In order to estimate extinction from NIR colors, we must consider the excess luminosity from accretion and warm circumstellar material. In $H$-band, excess luminosity or spectral veiling of 0.8 has been measured for T Tauri N by \citet{kasper02}. For T Tauri Sb, an $H$-band veiling of 0.7 and a $K$-band veiling of 2 were measured by \citet{duchene05}. Photospheric features in NIR spectra of T Tauri Sa have not yet been detected, preventing us from estimating the excess emission. We can, however, already say that it must be very high. 

It is widely believed that T Tauri Sa is extincted by an almost edge-on, small (3-5 AU radius), circumstellar disc hiding the stellar photosphere \citep[e.g.][]{koresko00,kasper02,beck04}. There is also an apparent lack of cold dust emission from T Tauri S, implying a rather small mass of its circumstellar material \citep{hogerheijde97}. The circumstellar disk around T Tauri Sa seems to be oriented north-south, while the circumstellar disk around T Tauri Sb and the Sa/Sb circumbinary disk are not far from face-on, approximately co-planar with the binary's orbit \citep{ratzka09}.

The near environment of T Tauri is also a source of surprisingly strong emission of molecular hydrogen. The spatial distribution of the H$_2$ emission displays a very complex pattern of multiple outflows on all observed angular scales \citep{herbst97,herbst07,saucedo03,gustafsson10}. Some of the H$_2$ emission is most likely generated by shock heating \citep{herbst97}, but \citet{saucedo03} also found fluorescent H$_2$ pumped by Ly$\alpha$, suggesting the action of low-density, wide opening-angle outflows driving cavities into the molecular medium. No significant H$_2$ emission is associated with the stars themselves \citep{kasper02,duchene05,herbst07}. 

The observed structures can be identified with several outflow systems. Originally, it was suggested that T Tauri N fuels the jet pointing west and ending in the Herbig-Haro object HH\,155 \citep{buehrke86}, leaving the Southeast-Northwest outflow to the hitherto unresolved T Tauri S. Using ground-based adaptive optics Fabry-Perot H$_2$ imaging, \citet{herbst07} found evidence that the East-West outflow is instead triggered by either T Tauri Sa or Sb. This in turn would mean that the Southeast-Northwest outflow would have to be attributed either to T Tauri N or to the southern component not triggering the East-West outflow. The presumably north-south orientation of T Tauri Sa's edge-on disk \citep{ratzka09} suggests that the East-West outflow indeed originates from T Tauri Sa.

Additional stellar components have repeatedly been reported in the literature. While \citet{nisenson85} detected a previously unknown source 0$\farcs$27 north of T Tauri N, \citet{ray97} reported the presence of a source roughly 0$\farcs$3 south of T Tauri N. Finally, \citet{csepany15} detected a feature indicative of a new companion 144\,mas south of T Tauri N. So far, none of these claims have been confirmed by follow-up observations.


\section{Observations and data reduction}
We observed T Tauri with SPHERE \citep{beuzit08, kasper12} on 9 December 2014 within the ESO Science Verification Program 60.A-9363 (PI M. Kasper) using the InfraRed Dual-band Imager and Spectrograph \citep[IRDIS,][]{dohlen08} in classical imaging mode, and on 23 January 2015 in Program 60.A-9364 (PI G. Cs\'ep\'any) using the IRDIFS extended (IRDIFS-EXT) mode. In the latter case, SPHERE observes simultaneously with the Integral Field Spectrograph \citep[IFS,][]{claudi08} in $J$- and $H$-band (R$\sim$30), and with IRDIS in two $K$-band filters optimized for the spectral differential detection of extra-solar planets \citep[IRDIS DBI,][]{vigan10}. Both observations were carried out in field-stabilized mode with (for IRDIFS-EXT) and without (for IRDIS classical imaging) the apodized Lyot coronagraph (ALC). The ALC is optimized for observations from $Y$- to $H$-band providing an inner working angle defined by the mask diameter of 185\,mas. The extreme adaptive optics system SAXO \citep{fusco14} corrected for atmospheric turbulence.

The classical imaging data were acquired in different narrow-band filters: Cont$J$ (1211\,nm), Pa$\beta$ (1282\,nm), H$_2$ (2122\,nm), Br$\gamma$ (2167\,nm), and Cont$K2$ (2267\,nm). For the astrometry and photometry, 160 two-second exposures were recorded in each filter using a neutral density filter with factor ten attenuation (ND1) to avoid point spread function (PSF) saturation. The SPHERE webpage\footnote{http://www.eso.org/sci/facilities/paranal/instruments/sphere.html} provides the filter transmission curves. For deep imaging in the $\nu$ = 1-0 S(1) line of H$_2$ and the adjacent continuum, 80 eight-second exposures were recorded in each filter without a neutral density filter. This led to saturation of the PSF core of T Tauri N in all filters and of T Tauri Sa in the $K$-band filters. Two hundred sixteen-second exposures were recorded with the IFS. These data are not saturated, because T Tauri N was masked by the coronagraph, and T Tauri Sa is considerably fainter in $J$- and $H$-band than in the $K$-band.

Calibration data (sky, flat field, etc.) were recorded as part of the standard daily calibration. SPHERE frequently observes astrometric calibration fields to monitor plate scale and field orientation. For the coronagraphic observations, we recorded images with T Tauri N shifted away from behind the Lyot mask for flux calibration. In this case, the ND1 filter was inserted to avoid PSF saturation. For the IRDIS classical imaging, we also recorded the instrumental PSF in all narrow-band filters using a SPHERE-internal calibration point source. This is required to calibrate the ghosts and filter defects which are unfortunately present in many of the SPHERE narrow-band filters. The IRDIS pixel size  $12.251\pm0.005$\,mas and the IFS image plate scale $7.46\pm0.02$\,mas per pixel are those provided in the SPHERE user manual.

We used the SPHERE data reduction pipeline \citep{pavlov08} to create backgrounds, bad pixel maps and flat fields. We reduced the raw data by subtracting the background, replacing bad pixels by the median of the nearest valid pixels, and finally dividing by the flat field. We also used the pipeline to create the IFS x-y-$\lambda$ data cube. Parts of this cube were collapsed along the wavelength axis to create broad-band images in $J$-band (1140-1350\,nm) and the short end of $H$-band (1490-1640\,nm. Note that a filter in SPHERE cuts off the long end of the $H$-band in front of the IFS, in order to limit sky background). We also created images in the wavelength bin 1400-1450\,nm, the shortest wavelength range in which T Tauri Sa could be seen.

The IRDIS images were centered with respect to each other using T Tauri Sb, which is never saturated and is readily visible in all filters, as a reference. For precise photometry and astrometry, we used the non-saturated PSF of T Tauri N to create a synthetic triple star system. The relative positions and magnitudes of the stars were determined by a downhill simplex nonlinear minimization of the quadratic residuals left after subtraction of the synthetic system from the T Tauri data. This data reduction strategy is vastly superior to simple aperture photometry in crowded areas with PSF overlap such as T Tauri. The accuracy is limited by systematic errors such as flat field uncertainties, which we assume to be of the order of one percent. We created images of spatially extended emission from circumstellar material by subtracting the synthetic image of the T Tauri triple from the saturated-star frames; otherwise the difference images would be dominated by PSF residuals close to the stars. 

Classical photometry, using small apertures with the diameter of the Airy disk, was applied to the IFS data for T Tauri Sa and Sb, because T Tauri N's PSF is strongly altered by the coronagraph and can no longer be used for the simultaneous PSF fitting described above. At 1.425\,$\mu$m, the background at the position of the fainter Sa was estimated from shorter wavelength data (1.375\,$\mu$m), where Sa is no longer seen. In $H$-band, we estimated the background from an equivalent patch on the opposite side of Sb. The errors for the classical aperture photometry are significantly higher than for the PSF fitting, up to $\sim$0.5 magnitudes for T Tauri Sa at 1.425\,$\mu$m.

\section{Results} 

\subsection{Continuum imaging} 

\begin{figure}
   \centering
   \includegraphics[width=\hsize]{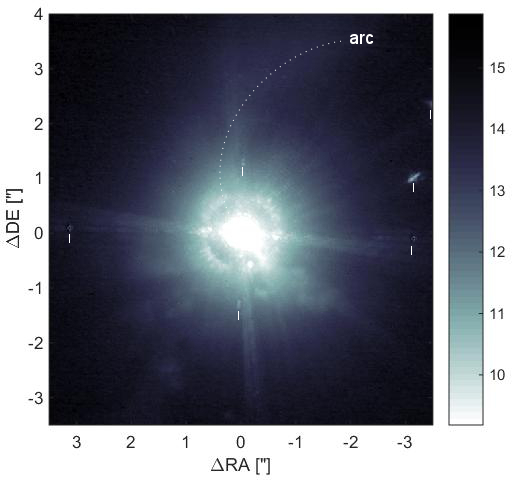}
      \caption{T Tauri in $J$-band extended emission. The image is displayed with a logarithmic color scale showing the flux level in magnitudes per arsecond$^2$. The dotted arc indicates the northern arm of the reflection nebulosity discussed by \citet{stapelfeldt98}. The image also contains several artifacts created by the narrow-band filters (above the vertical line segments), and by the AO correction (see text).
              }
         \label{fig1}
   \end{figure}

\begin{figure}
   \centering
   \includegraphics[width=\hsize]{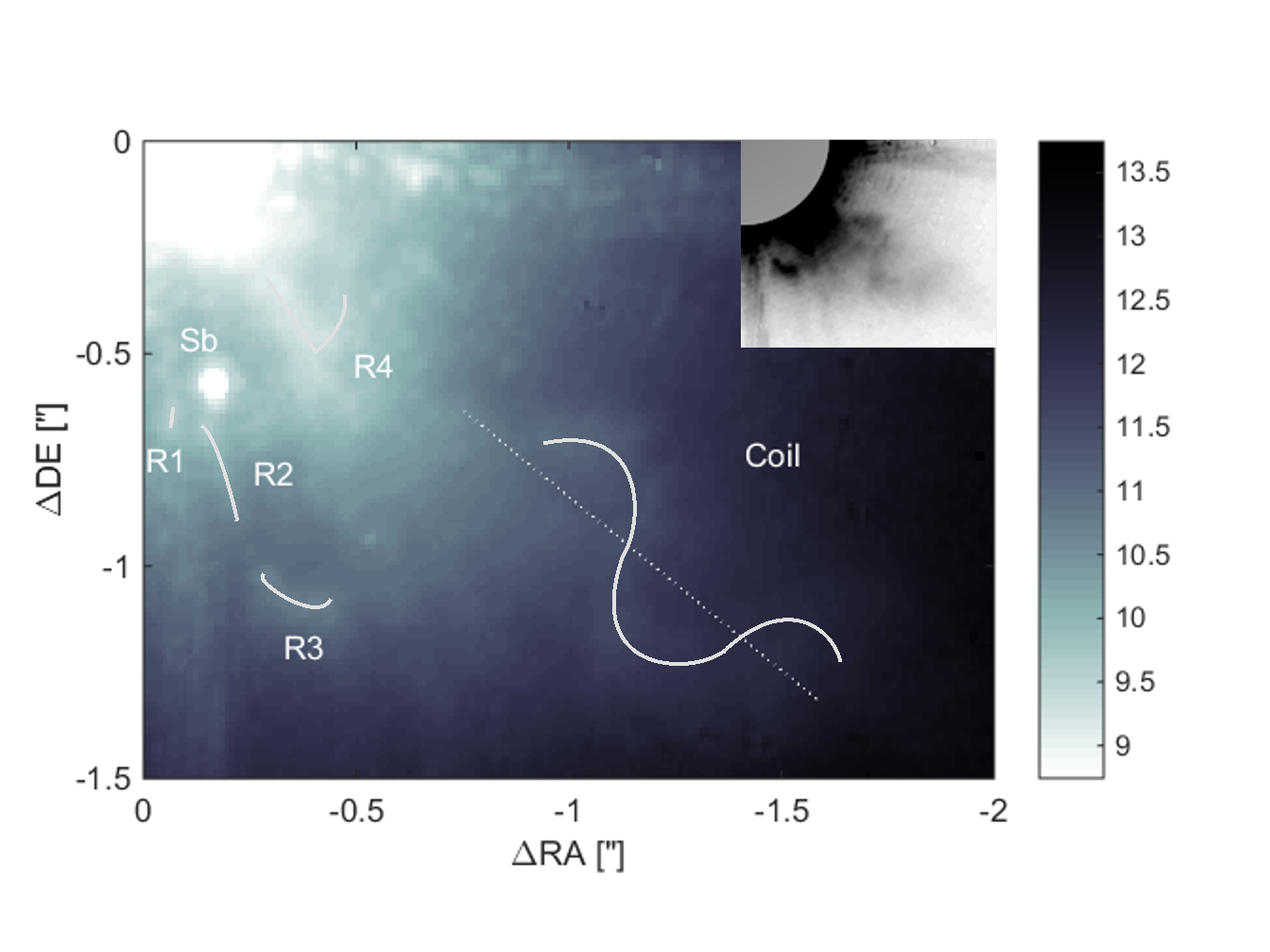}
      \caption{Zoom on the area southwest of T Tauri N of Figure\,\ref{fig1} showing the coiling structure and the new reflection nebulosity features R1-R4 in the vicinity of T Tauri S. The dashed line connects the inflection points of the coil and points back to  the vicinity of  T Tauri N. The image tile in the upper right corner shows the same area after subtracting a radial fitted exponential in the outer part of the image to visually enhance the contrast of R3 and the coil.
							}
         \label{fig2}
   \end{figure}

Figure\,\ref{fig1} shows an IRDIS $J$-band image composed of the sum of the Pa$\beta$ and CntJ narrow band images, and Figure\,\ref{fig2} zooms in on the area southwest of T Tauri N, which is centered on the upper leftmost pixel. 

\begin{figure}
   \centering
   \includegraphics[width=\hsize]{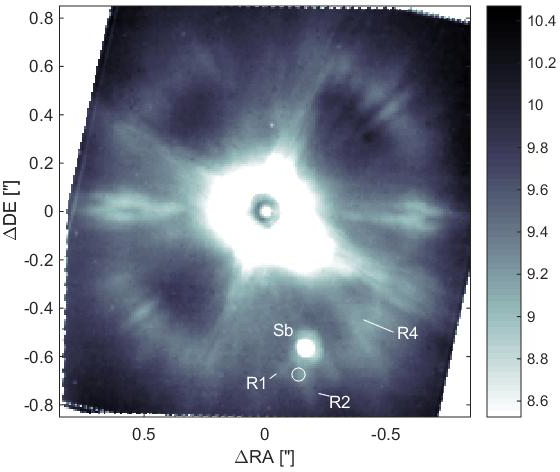}
      \caption{Coronagraphic $J$-band (1140-1350\,nm) image obtained with the IFS displayed in a logarithmic color scale showing the flux level in magnitude per arsecond$^2$. The small circle indicates the position of T Tauri Sa as seen at longer wavelengths.}
         \label{fig3}
   \end{figure}

\begin{figure}
   \centering
   \includegraphics[width=\hsize]{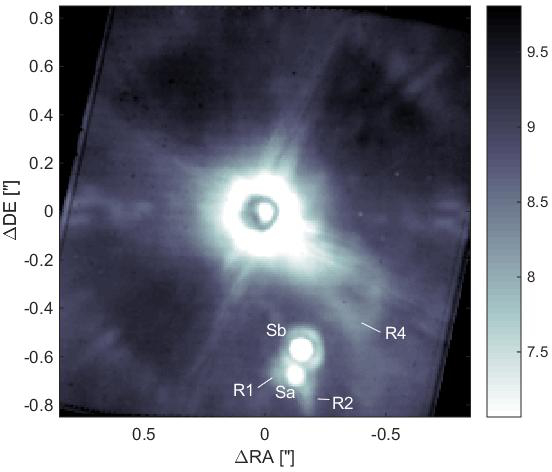}
      \caption{Coronagraphic $H$-band (1490-1640\,nm) image obtained with the IFS displayed in a logarithmic color scale showing the flux level in magnitude per arsecond$^2$.
              }
         \label{fig4}
   \end{figure}

\begin{figure}
   \centering
   \includegraphics[width=\hsize]{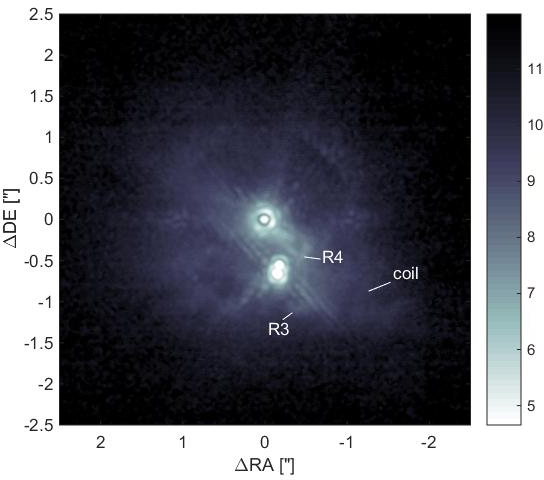}
      \caption{Coronagraphic $K1$-band image of T Tauri displayed in a logarithmic color scale showing the flux level in magnitude per arsecond$^2$. T Tauri N is masked by the ALC coronagraph. 
              }
         \label{fig5}
   \end{figure}

Figures\,\ref{fig3} (IFS $J$-band image, 1140-1350\,nm), \,\ref{fig4} (IFS $H$-band image, 1490-1640\,nm) and \ref{fig5} (IRDIS $K1$-band image) have been created from the IRDIFS-EXT observations. Here, a large fraction of the light from T Tauri N, the optically bright guide star for the adaptive optics system, is blocked by the ALC. The field of view of the reconstructed IFS cubes is 1$\farcs$7 across. In IRDIFS-EXT, the reconstructed x-y-$\lambda$ data cube consists of 39 $\lambda$-slices between 0.95\,$\mu$m and 1.68\,$\mu$m spaced by $\Delta\lambda=19.1$\,nm. Slices of this data cube can be combined to create images at the desired wavelength and spectral bandwidth. 

Some of the image structure in these figures are produced by the adaptive optics (AO) point-spread function (PSF). The large, diffuse ring-like structure centered on T Tauri N with approximate radii between 0$\farcs$6 ($J$-band) and 1\arcsec{} ($K$-band), for example, corresponds to the so-called control radius of the deformable mirror (DM). Beyond this radius, the DM with its finite number of actuators can no longer reproduce and correct for optical aberrations. The correction radius scales with wavelength, so it is almost twice larger in $K$-band than in $J$-band. The cross-like structure centered on T Tauri N are light diffracted by the telescope's secondary mirror support spiders. 

In the large field of view Figure\,\ref{fig1}, we detect an arc of reflection nebulosity with an approximate symmetry axis toward the west-northwest. The northern arm of this arc is traced by the dotted line. This arc was first imaged with the HST in the F555W, F675W, and F814W filters by \citet{stapelfeldt98}, who interpreted it as scattered light within an illuminated, axis-symmetric outflow cavity in a circumbinary envelope, viewed $\sim$45\degr{} from the outflow axis. It is readily seen in the $J$-band image, but we can no longer detect it in $K$-band (Figure\,\ref{fig5}).

Also very prominent is a new structure stretching to the southwest of T Tauri N and resembling a coil. We trace it outwards to about 2$\farcs$5 or 370\,AU projected on the sky at the distance of T Tauri. The coil is most easily seen in $J$-band, but is also detected in $K$-band. The dashed line in Figure\,\ref{fig2} connects the inflection points of the coil, and points back to the vicinity of T Tauri N at a position angle of $\sim$230\degr{}. 

The images further show several previously undetected features in NIR continuum emission, strongly suggesting that we are seeing reflection nebulosity. In Figure\,\ref{fig2}, we label these reflected light features R1-4 to differentiate them from the H$_2$ emission features reported by \citet{herbst07} and \citet{gustafsson10}. The inner $\sim$0$\farcs$65 show a feature, labeled R4, which resembles a short, rather straight line ending in a bow. It is most easily seen at longer wavelengths in the $H$- and $K$-bands (see Figures\,\ref{fig4} and \ref{fig5}), and appears near the location of the H$_2$ features 4, 5 and 6 of \citet{gustafsson10} and C2 of \citet{herbst07}.

In $J$- and $H$-band, the structure R2 resembles a $\sim$250\,mas long line pointing to the south from the current position of T Tauri Sb. The morphology of R2 changes significantly in $K$-band, where we can no longer detect the south-pointing line, but see a small structure of extended emission already reported by \citet{csepany15}. After a small gap of $\sim$150\,mas without a detection of significant emission, R2 appears to pass into the more prominent structure R3 roughly half an arcsecond south of T Tauri S, which curves away toward the west. R3 appears near the location of the H$_2$ feature 3 reported by \citet{gustafsson10}.

In $J$-band, we also detect a small knot of spatially unresolved emission (R1) about 80\,mas east of R2. This faint feature is seen with both instruments, IRDIS (Figure\,\ref{fig2}) and IFS (Figure\,\ref{fig3}). In $H$-band (Figure\,\ref{fig4}), R1 appears as a flux enhancement within the first Airy ring of the PSF of T Tauri Sa. Note that the similar region of the PSF of T Tauri Sb does not show this enhancement. 

While point-like emission from T Tauri Sa is readily detected in $H$-band, the star is no longer seen in the $J$-band at the position where it appears at longer wavelengths (indicated by the white circle in Figure\,\ref{fig3}). The shortest wavelength at which we can still see emission from T Tauri Sa itself is at 1.425\,$\mu$m.

\begin{figure}
   \centering
   \includegraphics[width=\hsize]{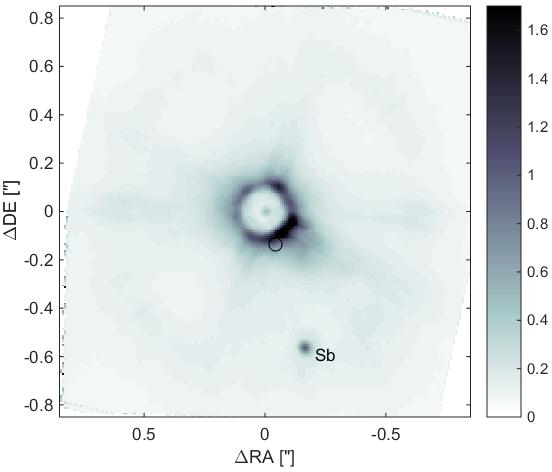}
      \caption{Same as Figure\,\ref{fig3}, but with a linear grayscale stretch normalized to the intensity of T Tauri Sb adjusted for visualization of the tentative companion candidate (CC, position indicated by the circle). The tentative CC was reported to be $\sim$4 times brighter than T Tauri Sb by \citet{csepany15} using the same data-set.
										}
         \label{fig6}
   \end{figure}

\begin{figure}
   \centering
   \includegraphics[width=\hsize]{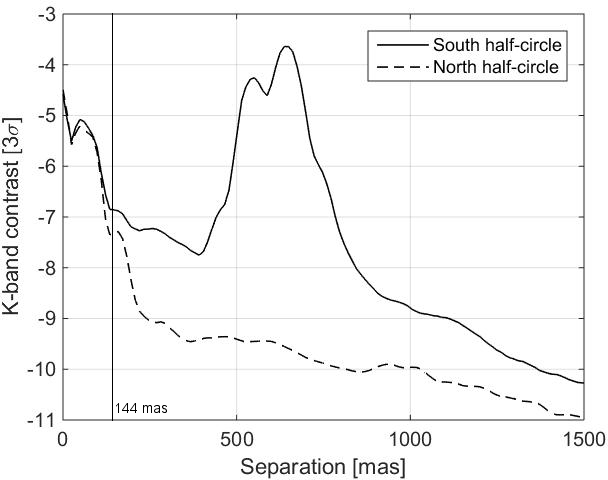}
      \caption{Magnitude contrast (3$\sigma$) obtained from the raw coronagraphic $K1$-band image (Figure\,\ref{fig5}) with respect to T Tauri N. The contrast is derived as the flux standard deviation calculated over 4-pixel wide annuli at a given separation. We distinguish between the North and South half-circles, because T Tauri S and its PSF residuals limit the contrast to the south. The thin vertical line indicates the separation of the tentative companion candidate at 144\,mas \citep{csepany15}.
										}
         \label{fig7}
   \end{figure}

\citet{csepany15} reported a tentative companion candidate (CC) to T Tauri N at separation 144\,mas and position angle 198\degr{} at a magnitude contrast of $\Delta J \sim 4.4$. Given that T Tauri Sb would be $\sim$1.5 magnitudes fainter than the tentative CC, our $J$-band reduction of the same data set excludes its presence as illustrated in Figure\,\ref{fig6}. Also the $K$-band imaging data (Figure\,\ref{fig5}) does not show the CC, despite the achieved contrast of about 7 magnitudes (see Figure\,\ref{fig7}, southern half circle). As $(J-K) \sim 1.4$ for T Tauri N \citep{herbst07}, and even the bluest brown dwarfs have colors $(J-K) > -1$, the CC should appear at $\Delta K < 6.8$ and would have readily been detected also in our $K$-band imagery. We conclude that the tentative CC is not real and probably was a data reduction artifact.

\subsection{Molecular hydrogen line emission} 

   \begin{figure}
   \centering
   \includegraphics[width=\hsize]{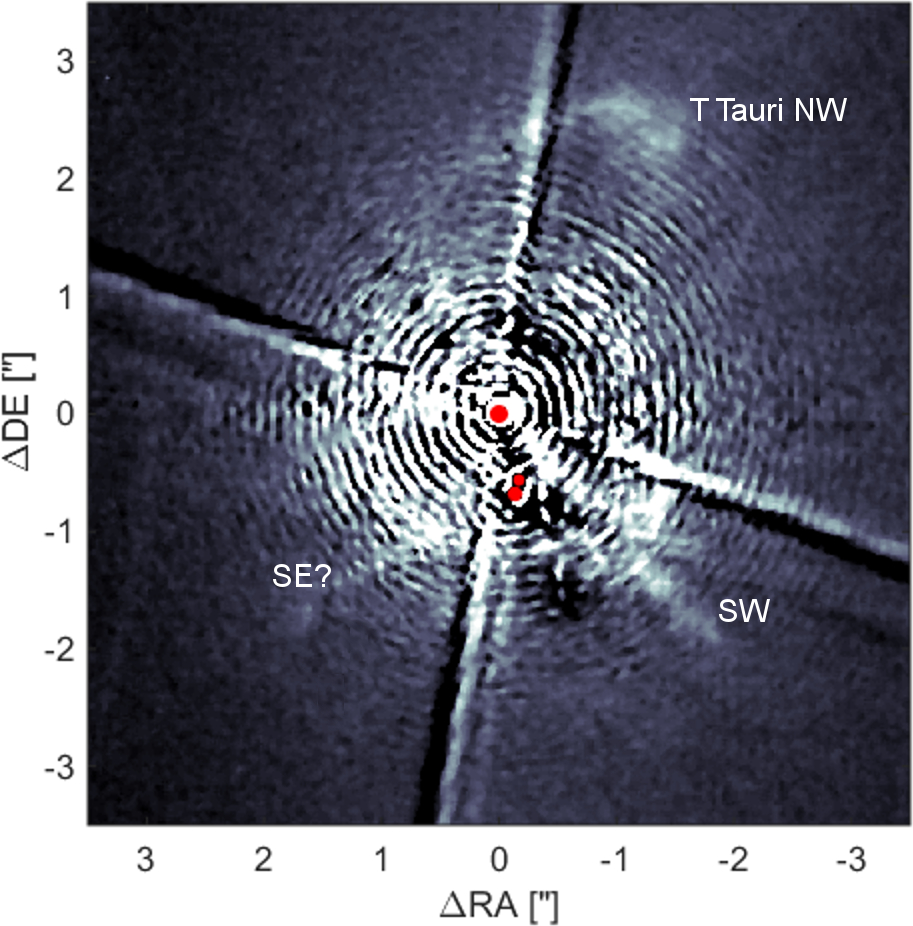}
      \caption{H$_2$ line emission of T Tauri. The stars are indicated by the red circles.The arc/ring-like artifacts are due to the mismatch in wavelength between the two narrow band images and made prominent by the grayscale stretch adjusted to show the low surface brightness H$_2$ features.
              }
         \label{fig8}
   \end{figure}

Figure\,\ref{fig8} shows H$_2$ line emission from the close vicinity around T Tauri. The image was created by subtracting the Br$\gamma$ image from the H$_2$ image. Br$\gamma$ emission originates from the immediate vicinity of the stars and is not spatially resolved in published imagery nor in our data, so it is a good proxy for a continuum filter image next to H$_2$. As the Airy pattern scales with wavelength, even the small difference of 45 nm between both narrow-band filters leaves significant Airy ring residuals after the subtraction. Also, the spectral width of the filters of 31 nm full width at half maximum is much broader than the intrinsic line widths of less than one nm \citep{duchene05}, thereby transmitting continuum flux and reducing the sensitivity. Finally, the observations in the two filters are separated in time by about ten minutes, during which time the telescope pupil rotated with respect to the sky. Therefore, the cross-like structure with T Tauri N in the center, produced by telescope M2 support structure diffraction residuals, rotated as well and does not subtract out. These shortcomings lead to higher image residuals than achievable with a high spectral resolution IFS or Fabry-Perot interferometer, and do not allow us to obtain a high SNR H$_2$ image in the immediate vicinity of the stars similar to those presented by \citet{gustafsson10}. Nevertheless, the SPHERE narrow filter imaging data clearly show spatially resolved H$_2$ in the vicinity of T Tauri. 

We readily identify the well-known H$_2$ region T Tauri NW discussed by \citet[e.g.][]{herbst96,herbst07} in our data, which has moved further to the northwest with respect to the stars when compared to its position of 2002 \citep{herbst07}. The apex of T Tauri NW is now located at $\sim$2$\farcs$8 and position angle $\sim$328\degr{} from T Tauri N.

Figure\,\ref{fig8} also shows a knot of H$_2$ emission about 1$\farcs$4 west and 1$\farcs$5 south of T Tauri N. Its position is similar to that of feature U reported by \citet{herbst07}. At a position angle of $223 \pm 1$\degr, this SW knot is also near to the coiling structure seen in reflected light (Figure\,\ref{fig1}), but the features do not overlap. The H$_2$ flux density of the SW knot is similar to that of T Tauri NW, and it extends somewhat further to the SW at lower intensity. There may be another similarly extended feature pointing from T Tauri S to the southeast at a PA of $119 \pm 1$\degr{}, but the signal-to-noise ratio is too low to say this with certainty.

\section{Analysis and Discussion} 

\subsection{The South--Western coiling structure and feature R4} 

Our data reveal a coiling structure to the southwest of T Tauri N, which can be traced from $\sim$1\arcsec{} out to $\sim$2$\farcs$5, or from 150 to 370\,AU projected on the sky. It is apparently reflected light seen at NIR wavelengths and should correspond to areas of enhanced density of ambient matter illuminated by the stars. A line drawn through the inflection points of the coil points to the vicinity of T Tauri N. The spatial length of one period of the coil is $\sim$0$\farcs$73 or $\sim$108 AU projected on-sky.

The coiling structure is reminiscent of that produced by the precessing jet of the pre-planetary nebula IRAS 16342-3814 \citep{sahai05}. In their paper, the authors argue that the jet beam is not seen directly, but via its interaction with the ambient circumstellar medium, i.e., imprinting on the expanding shell, which produces a compressed structure of enhanced density in the shape of a corkscrew. Generally, a precessing outflow could be the consequence of a binary system, caused either by orbital motion of the outflow's source or by outflow precession due to tidal effects. In the first case, the precession period would be the binary's orbital period, while it would be much longer in the latter case. 

Interferometic obervations of T Tauri N with a spatial resolution of a couple of millarseconds \citep{akeson02} exclude the presence of a stellar companion to T Tauri N with an orbital separation greater than a few tenths of an AU. The orbital period of a binary tighter than this would be measured in tens of days. Interestingly, \citet{ismailov10} found periodic variation in spectral features of T Tauri N on a time scale of 33 days, which are not yet understood. In order to move by one cycle of the coil, or $\sim$108 AU, in some tens of days, the interstellar medium would have to expand at a velocity of several thousand km/s, which is an order of magnitude higher than typical outflow velocities. So we can exclude that orbital motion of a tight binary T Tauri N is the origin of the coiling structure. 

The actual oscillation period can be crudely constrained from the time interval required by the compressed material to traverse one period of the coiling structure, i.e., $\sim$108 AU, projected on-sky. Assuming that we see the structure moving in the same direction as the spatially nearest H$_2$ features 5 and 6 of \citet{gustafsson10} at an inclination $\sim$20\degr{}, only about one third of the motion is in the plane of the sky. Making further the assumption that the compressed material expands at a velocity $\sim$50-100 km/s, a typical value for outflows from T Tauri stars, one would see the coils moving by $\sim$3.5-7\,AU or $\sim$25-50\,mas per year. The resulting precession period would be of the order of 15-30 years. At larger inclination angles, if the outflow would be close to the plane of the sky, the projected motion could be up to three times larger, and the precession period up to three times shorter. 

A period of 15-30 years could still be consistent with axial precession within a tight binary T Tauri N. Young stars are rapid rotators and expected to show significant flattening. T Tauri N itself has a rotation period of 2.8 days \citep{herbst86}, which is one order of magnitude shorter than that of the Sun. As the precession period is inversely proportional to orbital distance cubed, and scales linearly with the flattening \citep[e.g.][]{williams94}, axial precession rates of the order of a few tens of years can occur in a young and tight binary system. 

Another explanation for the coiling structure could be a jet-like outflow launched from within the T Tauri S system and imprinting on the cavity carved by another outflow, e.g., the cavity of the northwest outflow from T Tauri N, which has been proposed by \citet{stapelfeldt98}. In this case, the orbital period for the T Tauri Sa-Sb binary of 27$\pm$2 years \citep{koehler15} would be in a nice agreement with the predicted period of the coils.

Closer in toward T Tauri N, the coiling structure appears to connect to the reflection nebulosity feature R4 (most prominent in the $H$- and $K$-band Figures\,\ref{fig4} and \ref{fig5}). R4 is a rather straight line pointing from T Tauri N to the southwest, where it merges at $\sim$630\,mas from T Tauri N (or west-northwest of the current position of T Tauri Sb) into a bow. The straight part is too short to exclude that it actually has a curvature similar to that observed in the coiling structure. R4 is seen near the location where \citet{duchene05} and \citet[][features 5 and 6]{gustafsson10} detected blue-shifted H$_2$ emission, and where \citet{herbst07} placed the feature C2. \citet{gustafsson10} found no proper motion of these features during a time-span of two years and suggested that the flow is hence moving mostly towards the observer and seen at the low inclination ($\sim$20\degr{}) of T Tauri N. The bow morphology of R4 pointing toward T Tauri N provides further evidence that the features observed in the area are indeed produced by the T Tauri N outflow.

%

\subsection{Extended emission in the vicinity of T Tauri S} 

Using NACO imaging data from 2001 to 2009, \citet{vanBoekel10} discovered a ring of $K$-band continuum emission around T Tauri Sa with intensity of $\sim$5\% relative to the peak of T Tauri Sb. The authors suggest that this emission traces an outflow cavity with an inner area devoid of material. In our images, we identify continuum emission in the vicinity of T Tauri S in the $J$-, $H$- and $K$-bands. The SPHERE $K$-band data (Figure\,\ref{fig5}) do not show the ring structure, but a hook-like feature of extended emission $\sim$230\,mas south of the current position of T Tauri Sb \citep{csepany15}. This difference in appearance could be caused by the dynamical evolution of the close-in environment of the 27-year orbit binary system T Tauri Sa/Sb. In the shorter wavelength $J$- and $H$-bands, the emission resembles an almost straight line extending to the south of T Tauri Sb (feature R2 in Figures \ref{fig2} and \ref{fig4}). We measure intensities along R2 of $\sim$1.5\%, $\sim$2\%, and $\sim$3\%  at $J$-, $H$-, and $K$-band, respectively, relative to T Tauri Sb.

Our images also reveal an arc-like feature R3 at about 550 mas to the south-southwest of T Tauri Sb (most prominent in Figure \ref{fig2}). In contrast to R2, feature R3 appears bluer than T Tauri Sb with maximum pixel intensities of $\sim$1.5\% in $J$-band and $\sim$0.15\% in $K$-band (we do not have an $H$-band image, because R3 is outside the IFS field of view) relative to T Tauri Sb. R3 appears to be a continuation of R2 and could be part of the same outflow system consistent with blue-shifted H$_2$ emission found by \citet[][feature 3]{gustafsson10} and \citet[][C3]{herbst07}. Proper motion and radial velocity data let \citet{gustafsson10} measure an inclination $\sim$60\degr{} for this outflow, which they assigned to the T Tauri S binary with a preference for T Tauri Sb. The appearance of the R2/R3 reflection nebulosity features are consistent with this model, in which R2/R3 could trace the cavity walls of the T Tauri Sb outflow. The rather short length of R2 of around 200 mas or 30 AU could also explain why we do not see coiling or curvature due to orbital motion in the T Tauri Sa/Sb binary. The flow is traveling at $\sim$32 km/s \citep{gustafsson10} on the sky, so it would need just 4-5 years to cover 30 AU, while T Tauri Sb is on its orbit west of T Tauri Sa since more than 15 years.

Finally, Figures\,\ref{fig2} and \ref{fig3} show a faint spatially unresolved continuum emission clump, labeled R1, about 70\,mas east of T Tauri Sa. R1 is best seen in $J$-band where T Tauri Sa is no longer visible. At a projected distance of about 10 AU, it cannot be part of the T Tauri Sa circumstellar disk, which must be considerably smaller, given the semi-major axis of the T Tauri S binary of just 12.5 AU \citep{koehler15}. It could, however, be part of the T Tauri Sa outflow cavity suggested by \citet{vanBoekel10}.

\subsection{Molecular hydrogen emission, T Tauri Northwest and the Southwest outflow} 

H$_2$ line emission arises in a number of physical processes, the most common of which are shock excitation and UV fluorescence. T Tauri is an interesting case in which both mechanisms occur \citep{vanLangevelde94}. The flux ratios between the 1-0 S(1) and 2-1 S(1) lines can help to discriminate between the two, with higher ratios favoring shock excitation \citep{black76}. \citet{herbst96} determined a ratio of $>$15 for T Tauri NW, which argues strongly for shock excitation as the dominant mechanism. In addition, its classical Herbig-Haro bow shock morphology unambiguously suggests a terminal shock arising from a stellar outflow \citep{herbst07}.  

\begin{figure}
   \centering
   \includegraphics[width=\hsize]{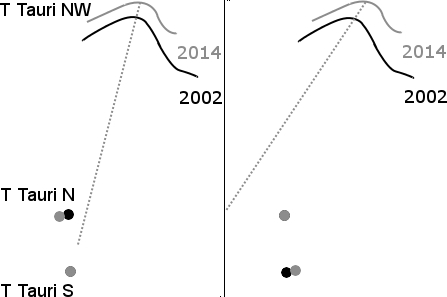}
      \caption{Overlay of the relative positions of T Tauri NW, N, and Sa in 2002 \citep{herbst07} and 2014 (our data). The left panel shows the result when using T Tauri Sa as the centring reference, while the right panel uses T Tauri N.
              }
         \label{fig9}
   \end{figure}

Comparing our new 2014 data to the 2002 NACO image of T Tauri NW \citep{herbst07}, we can attempt to estimate which star triggered the corresponding outflow. For this, we first matched the plate scales of the NACO and SPHERE images. Then, we created two sketches (one for 2002 and one for 2014) by drawing circles centered on T Tauri N and Sa and by drawing the outline of T Tauri NW by eye. From these, we finally created two overlays shown in Figure\,\ref{fig9}. In the overlay shown on the left, the 2002 and 2014 sketches are centered with respect to T Tauri Sa, while the overlay on the right shows the situation in the rest frame of T Tauri N.

In the launching star's rest frame, the motion of T Tauri NW would point straight away from the star, if accelerated orbital motion can be neglected. This is a reasonable approximation as long as the time since the ejection of the T Tauri NW outflow is small compared to the orbital period of the T Tauri N/S binary of $4200^{+5000}_{-3400}$ years \citep{koehler15}. Figure\,\ref{fig9} then strongly suggests that the T Tauri NW outflow is launched from within the T Tauri S system, because the motion of T Tauri NW points to T Tauri Sa in its rest frame (left panel). In the rest frame of T Tauri N (right panel), instead, the motion of T Tauri NW points to a position well east of the stars. The accuracy of this qualitative analysis is, however, not sufficiently high to differentiate between T Tauri Sa and Sb as a possible origin of the T Tauri NW outflow.

T Tauri NW has moved by about 200 mas at a position angle of $\sim$345\degr{} from T Tauri Sa between 2002 and 2014. This corresponds to an average velocity of  $\sim$17 mas/yr or $\sim$12 km/s perpendicular to the line of sight over the past 12 years. As the radial velocity of H$_2$ emission from T Tauri NW is quite small with values between $\sim$-3 km/s \citep{herbst97} and $\sim$-7 km/s \citep{gustafsson10} with respect to the stars, T Tauri NW moves at $\sim$12-14 km/s with an inclination angle of $\sim$60-75\degr{}. This is in good agreement with the SE-NW outflow's inclination of $\sim$70\degr{} derived by \citet{gustafsson10} in the area southeast of T Tauri S. The inclination is, however, not consistent with the nearly face-on orientation of T Tauri N \citep{herbst86, herbst97}, providing further evidence that the T Tauri S system is the source of the SE-NW outflow. 


Figure\,\ref{fig8} also shows an area of enhanced H$_2$ emission resembling a knot merging into a straight line pointing to the southwest of T Tauri N at a position angle of $223 \pm 1$\degr{}, similar to the direction of the coiling structure seen in reflected light (PA $\sim$230\degr{}). This H$_2$ emission could represent shocked material along a fast flow periphery where projected velocities are low enough to generate H$_2$ line emission efficiently as discussed by \citet{herbst97}. Our SW H$_2$ emission is an extension to the line formed with the clumps of H$_2$ emission observed in the area northwest and west of T Tauri S by \citet[e.g.][]{duchene05, herbst07}, which were rather stationary over several years and blue-shifted, i.e., moving towards the observer \citep{beck08, gustafsson10}. These features could therefore all belong to the same outflow launched by T Tauri N. In addition, the close spatial match of our detected H$_2$ emission to the southwest of T Tauri N with the coiling structure, suggests that it is associated with the same outflow as well.
  

\subsection{Photometry and extinction to T Tauri Sb} 

Table\,\ref{table:ttau_phot} summarizes the relative photometry of the stars in the T Tauri system in December 2014 and January 2015. We assume that T Tauri N (Spectral type early K) is not variable and has NIR magnitudes of $J=7.1$, $H=6.2$ and $K=5.7$ \citep{herbst07}. At the time of our observations, the apparent magnitudes of T Tauri Sa were then $J > 17.5$, $H \sim 11.8 $ and $K \sim 7.9$. Here, we use our SPHERE narrow-band and IFS data to interpolate to standard central wavelengths in the NIR bands of 1.25\,$\mu$m ($J$), 1.65\,$\mu$m ($H$) and 2.2\,$\mu$m ($K$). T Tauri Sa is hence entering another phase of elevated brightness, rising from $K \sim 10$ in 2005 and $K \sim 8.5$ during the period 2007-2010 \citep{vanBoekel10}. The shortest wavelength at which we can unambiguously detect emission of T Tauri Sa in the IFS data is 1.425\,$\mu$m. This is approximately half-way between the $J$- and $H$-bands. We adopt an apparent magnitude of $M_{1.425\mu m} = 6.6$ for T Tauri N and derive $M_{1.425\mu m} = 14.6$ for T Tauri Sa. 

Similarly, we measure apparent magnitudes of T Tauri Sb of $K \sim 8.9$, $H \sim 10.8$ and $J \sim 13.3$. T Tauri Sb is an M0.5 star, which typically have intrinsic $(J-K)$ colors of $\sim$0.85 \citep{newton14}. T Tauri Sb shows a spectral veiling of about two in the $K$-band \citep{duchene05}, i.e., the excess flux from warm circumstellar material is twice as bright as the photospheric emission. Hence, in the $K$-band, T Tauri Sb appears to be three times or 1.19\,mag brighter than the star really is. The $J$-band is generally considered to suffer the least from spectral veiling, because it maximizes the ratio between photospheric emission and the combined excess from hot accretion shock emission (bright at short wavelengths, UV) and circumstellar dust (bright at long wavelengths, $K$-band to mid IR) \citep{hartmann90}. We use the empirical relation of $K$- to $J$-band veiling by \citet{cieza05} to estimate a $J$-band veiling of 0.5 for T Tauri Sb, increasing the apparent magnitude of the star by 0.44\,mag. Subtracting i) the veiling contribution of $1.19 - 0.44 = 0.75$ mag and ii) the star's intrinsic $(J-K) \sim 0.85$ from the measured $(J-K) \sim 4.4$, we estimate the reddening of T Tauri Sb $E(J-K) = 2.8$. With extinction ratios $A_J/A_V = 0.26$ and $A_K/A_V = 0.09$ \citep[e.g.][]{cieza05}, we then derive the extinction in the direction of T Tauri Sb $A_K = 1.48$\,mag or $A_V = 16.5$\,mag, very similar to the previously measured $A_V = 15$\,mag \citep{duchene05}. This result provides further evidence that T Tauri Sb is subject to a roughly constant extinction along its orbit.


\begin{table*}
	\caption{Relative photometry of the T Tauri stars. T Tauri Sa is not detected short-wards of 1.425\,$\mu$m, so we provide magnitude limits only. Values at 1425\,nm and in $H_\text{IFS}$ (1490-1640\,nm) are derived from the IFS data, the others from IRDIS classical imaging.}
	\label{table:ttau_phot}
	\centering
	\begin{tabular}{l c c c}
		\hline\hline
		Filter & N-Sa	& N-Sb & Sb-Sa \\
    \hline
    Cont$J$ (1211\,nm) & $>10.7$  & $6.45\pm0.01$ & $>4.25$ \\
		Pa$\beta$ (1282\,nm) & $>10.1$ & $5.89\pm0.01$ & $>4.2$ \\
		1425\,nm & $8\pm0.5$ & $5.35\pm0.1$ & $2.65\pm0.5$ \\
		$H_\text{IFS}$ (1490-1640\,nm) & $6.15\pm0.2$ & $4.8\pm0.05$ & $1.35\pm0.2$ \\
		Br$\gamma$ (2167\,nm) & $2.37\pm0.01$ & $3.32\pm0.01$ & $-0.95\pm0.01$ \\
		Cont$K2$ (2267\,nm) & $2.05\pm0.01$ & $3.13\pm0.01$ & $-1.08\pm0.01$ \\
    \hline
  \end{tabular}
\end{table*}

\section{Summary and conclusions} 

T Tauri has always provided new surprises and new insights when an enhanced observational capability comes along. With our SPHERE narrow-band imaging and integral field spectroscopy data, we have taken another step toward understanding this enigmatic and archetypical young stellar object.

The newly-detected coiling structure to the southwest is a reflection nebulosity seen at NIR wavelengths, suggesting that it is ambient circumstellar medium compressed by a precessing jet in the shape of a corkscrew. The spatial period of the coils suggests a precession period of the order 15 to 30 years, provided that the outflow is seen at a relatively high inclination. We identify two mechanisms which could produce such a precession period: i) axial precession in a very tight (semi-major axis smaller than a few tenths of an AU) T Tauri N binary system, or ii) orbital motion of the T Tauri Sa/Sb binary whose period of 27 years is in good agreement with the expected time span. The present data, however, does not allow us to unambiguously identify the star launching the coil-producing jet. 

Towards T Tauri N, the coiling structure appears to connect to the reflection nebulosity R4, which resembles a straight line merging into a bow. The orientation of the bow provides evidence that we see the effect of the T Tauri N outflow. This conclusion is supported by the location of R4 near previously detected H$_2$ emission features, which are seen at relatively low inclination to the line of sight consistent with an outflow from T Tauri N. Near to the coiling structure in the southwest, we detect a knot of H$_2$ emission merging into a straight line, which could represent shocked material along a fast flow periphery where projected velocities are low enough to generate H$_2$ line emission efficiently. 

We also detect new reflection nebulosity just south of T Tauri S. The two features, a relatively straight line pointing south from T Tauri Sb (R2) and an arc-like structure somewhat further south (R3), could both trace the cavity of the moderately inclined SW outflow from T Tauri Sb suggested by \citet{gustafsson10}. 

The well known H$_2$ emitting region T Tauri NW is readily detected in our narrow-band data. The comparison of the positions of T Tauri NW with respect to T Tauri N and S between 2002 \citep{herbst07} and our 2014 data, provides further evidence that the Southeast-Northwest outflow triggering T Tauri NW is likely to be associated with T Tauri S. Over the past decade, it has moved with average velocity $\sim$12 km/s perpendicular to the line of sight at a position angle of 345\degr{} with respect to T Tauri S.

We measure an extinction $A_V = 16.5$\,mag towards T Tauri Sb, consistent with previous estimates, and providing further evidence for a nearly constant extinction along its orbit. Our analysis also rejects the recently proposed tentative companion candidate to T Tauri N and sets tight constraints for additional stellar companions in the vicinity of T Tauri N at separations larger than about 0$\farcs$1 (15 AU). If, however, the coiling structure is indeed produced by axial precession due to tidal effects in a close binary T Tauri N, there could remain at least one more star to be discovered in the T Tauri system.


%


\begin{acknowledgements}
This work is based on observations performed with VLT/SPHERE under program IDs 60.A-9363(A) and 60.A-9364(A). We would like to thank the astronomers and the instrument support team at the VLT for the observations in science verification.
\end{acknowledgements}


%
%

\bibliographystyle{aa}
\bibliography{kasperrefs}

\end{document}